\documentclass[12pt]{article}
\usepackage{graphics}
\usepackage{amssymb}

\newcommand{\e}{\mathrm{e}}

\newcommand{\N}{\mathbb{N}}
\newcommand{\R}{\mathbb{R}}
\newcommand{\Z}{\mathbb{Z}}
\newcommand{\B}{\mathcal{B}}
\newcommand{\D}{\mathcal{D}}
\newcommand{\J}{\mathcal{J}}

\newtheorem{claim}{Claim}[section]
\newtheorem{theorem}[claim]{Theorem}
\newtheorem{proposition}[claim]{Proposition}
\newtheorem{corollary}[claim]{Corollary}
\newtheorem{lemma}[claim]{Lemma}
\newtheorem{example}[claim]{Example}
\newtheorem{remark}[claim]{Remark}
\newtheorem{remarks}[claim]{Remarks}
\newenvironment{proof}[1][Proof]{\textsl{#1.} }{\ \rule{0.5em}{0.5em}}
\begin{document}

\title{Strong-coupling asymptotic expansion for Schr\"odinger
operators with a singular interaction supported by a curve in
$\R^3$}
\author{P.~Exner$^{a,b}$ and S.~Kondej$^{c}$}
\date{}
\maketitle

\begin{quote}
{\small \em a) Nuclear Physics Institute, Academy of Sciences,
25068 \v Re\v z \\ \phantom{a) }near Prague, Czech Republic
\\
b) Doppler Institute, Czech Technical University,
B\v{r}ehov{\'a}~7, \\ \phantom{a) }11519 Prague, Czech Republic
\\
c) Institute of Physics, University of Zielona G\'{o}ra, ul.
Szafrana 4a, \\ \phantom{a) } 65246 Zielona G\'{o}ra, Poland
\\

\phantom{a) }\texttt{exner@ujf.cas.cz},
\texttt{skondej@if.uz.zgora.pl} }
\end{quote}

\begin{quote}
{\small {\bf Abstract.} We investigate a class of generalized
Schr\"{o}dinger operators in $L^2(\mathbb{R}^3)$ with a singular
interaction supported by a smooth curve $\Gamma$. We find a
strong-coupling asymptotic expansion of the discrete spectrum in
case when $\Gamma$ is a loop or an infinite bent curve which is
asymptotically straight. It is given in terms of an auxiliary
one-dimensional Schr\"{o}dinger operator with a potential
determined by the curvature of $\Gamma$. In the same way we obtain
an asymptotics of spectral bands for a periodic curve. In
particular, the spectrum is shown to have open gaps in this case
if $\Gamma$ is not a straight line and the singular interaction is
strong enough. }
\end{quote}


\section{Introduction}

The subject of this paper are asymptotic spectral
properties for several classes of generalized Schr\"{o}dinger
operators in $L^2(\mathbb{R}^3)$ with an attractive singular
interaction supported by a smooth curve or a family of such
curves. On a formal level, we can write such a Hamiltonian as
\begin{equation} \label{formal}
-\Delta -\tilde\alpha \delta (x-\Gamma )\,,
\end{equation}
however, a proper way to define the operator corresponding to the
formal expression is involved and will be explained in
Sec.~\ref{H} below\footnote{In particular, this is the reason why
we use here a formal coupling constant different from the
parameter $\alpha$ introduced in the condition (\ref{boucon})
below.}. A physical motivation for this model is to understand the
electron behavior in ``leaky'' quantum wires, i.e. a model of
these semiconductor structures which is realistic in the sense
that it takes into account the fact that the electron as a quantum
particle capable of tunelling can be found outside the wire --
cf.~\cite{EI} for a more detailed discussion.

One natural question is whether in case of a strong transverse
coupling properties of such a ``leaky'' wire will approach those
of an ideal wire of zero thickness, i.e. the model in which the
particle is confined to $\Gamma$ alone, and how the geometry of
the configuration manifold will be manifested at that. In the
two-dimensional case when $\Gamma$ is a planar curve this problem
was analyzed in \cite{EY1, EY2} where it was shown that apart of
the divergent term which describes the energy of coupling to the
curve, the spectrum coincides asymptotically with that of an
auxiliary one-dimensional Schr\"odinger operator with a
curvature-induced potential\footnote{A similar analysis was
performed in \cite{Ex} for smooth surfaces in $\mathbb{R}^3$ where
the asymptotic form of the spectrum is given by a suitable
``two-dimensional'' operator supported by the surface $\Gamma$.}.

The case of a curve in $\mathbb{R}^3$ which we are going to
discuss here is more complicated for several reasons. First of
all, the codimension of $\Gamma$ is two in this situation which
means that to define the Hamiltonian we cannot use the natural
quadratic form and have to employ generalized boundary conditions
instead. Furthermore, while the strategy of \cite{EY1, EY2} based
on bracketing bounds combined with the use of suitable curvilinear
coordinates in the vicinity of $\Gamma$ can be applied again, the
``straightening'' transformation we have to employ is more
involved here. Also the bound on the transverse part of the
estimating operators are less elementary in this case.

Let us review briefly the contents of the paper. We begin by
constructing a self-adjoint operator $H_{\alpha ,\Gamma }$ which
corresponds to the formal expression (\ref{formal}), where $\Gamma
$ is a curve in $\R^{3}$; this will be done in Sec.~\ref{Hamilt}.
To this aim we employ in the transverse plane to $\Gamma$ the
usual boundary conditions defining a two-dimensional point
interaction \cite[Sec.~I.5]{AGHH}. Recall that the latter is known
to have for any $\alpha\in\R$ a single negative eigenvalue which
equals $\xi _{\alpha }=-4\e^{2(-2\pi \alpha +\psi (1))}$, where
$-\psi (1)=0.5777...$ is the Euler constant. The main topic of
this paper are spectral properties of $H_{\alpha ,\Gamma }$ in the
strong-coupling asymptotic regime which means here that $-\alpha$
is large. The auxiliary operator mentioned above is given by
$$ S:=-\Delta -\frac{1}{4}\kappa
^{2}, $$
where $\Delta $ is the one-dimensional Laplace operator on the
segment parameterizing $\Gamma$ and $\kappa$ is the curvature of
$\Gamma $. Its discrete spectrum is non-empty unless $\Gamma$ is a
straight line; we denote the $j$-th eigenvalue as $\mu_j$. Our
main results can be then characterized briefly as follows:

\emph{Discrete spectrum:} If $\Gamma $ is a loop, we show in
Sec.~\ref{asloop} that the $j$-th eigenvalue $\lambda _{j}(\alpha
)$ of $H_{\alpha, \Gamma}$ admits an asymptotic expansion of the
following form,
$$
\lambda _{j}(\alpha )=\xi _{\alpha }+\mu _{j}+\mathcal{O}(e^{\pi
\alpha })\quad \mathrm{as} \quad \alpha \rightarrow -\infty\,,
$$ and the counting function $\alpha \mapsto \,\#\sigma
_{d}(H_{\alpha, \Gamma})$ satisfies in this limit the relation
$$
\#\sigma _{d}(H_{\alpha })=\frac{L}{\pi }(-\xi
_{a})^{1/2}(1+\mathcal{O} (e^{\pi \alpha })).
$$\
In addition, the last formula does not require $\Gamma $ to be a
closed curve as we shall show in Sec.~\ref{freeend}. Moreover, if
$\Gamma $ is infinite with $\kappa\ne 0$ and at the same time
asymptotically straight in an appropriate sense then the above
expansion for $\lambda _{j}(\alpha )$ holds again --
cf.~Sec.~\ref{infcur}.

\emph{Periodic curves} are discussed in Sec.~\ref{asperiod}; we
perform Bloch decomposition and use the same technique as above to
estimate the discrete spectrum of the fiber operators. In
particular, we find that if $\Gamma $ is periodic curve and
$\kappa (\cdot)$ is nonconstant then $\sigma(H_{\alpha ,\Gamma })$
contains open gaps for $-\alpha $ sufficiently large. In the
closing section we will show that the problem can be rephrased in
terms of a semiclassical approximation and list some open
problems.


\setcounter{equation}{0}
\section{Hamiltonians with curve-supported \\ perturbations}
\label{Hamiltsec} \setcounter{equation}{0}

\subsection{The curve geometry}

Let $\Gamma $ be a curve in $\R^{3}$ (either infinite or a closed
loop) which is assumed to be $C^{k},\, k\geq 4$. Without loss of
generality we may assume that it is parameterized by its arc
length, i.e. to identify $\Gamma$ with the graph of a function
$\gamma:\: I\to \R^{3}$, where $I=[0,L]$ (with the periodic
boundary conditions, $\gamma (0)=\gamma (L)$ and the same for the
derivatives) if $\Gamma$ is finite and $I=\R$ otherwise. One of
our tools will be a parametrization of some neighbourhoods of
$\Gamma$. To describe it let us suppose first that the curve
possesses the global Frenet's frame, i.e. the triple
$(t(s),b(s),n(s))$ of tangent, binormal, and normal vectors which
are by assumption $C^{k-2}$ smooth functions of $s\in I$; recall
that this is true if the second derivative of $\Gamma$ vanishes
nowhere.

The mentioned neighbourhoods are open tubes of a fixed radius
centred at $\Gamma$: given $d>0$ we call $\Omega _{d}:= \{x\in\R^3
:\: \mathrm{dist} (x,\Gamma)<d \}$. We will impose another
restriction on the class of curves excluding those with
self-intersections and ``near-intersections'', i.e. we suppose
that
  \begin{description}
  \item{(a$\Gamma 1$)}
there exists $d>0$ such that the tube $\Omega _{d}$ does not
intersect itself.
 \end{description}
Our aim is to describe $\Omega _{d}$ by means of curvilinear
coordinates, i.e. to write it as the image of a straight cylinder
$B_{d}:=\{ r\in [0,d),\, \theta \in [0,2\pi) \}$ by a suitable
map. If the Frenet frame exists we choose the latter as $\phi
_{d}:\:\D_{d}\to\R^3$ defined by
\begin{equation} \label{Tangdif}
\phi _{d}(s,r,\theta)= \gamma (s)-r \left\lbrack n(s)
\cos(\theta\!-\!\beta(s))+b(s)\sin(\theta\!-\!\beta(s))
\right\rbrack \,,
\end{equation}
where $\D_{d}:= I\times B_{d}$ and the function $\beta $ will be
specified further. For convenience we will denote the curvilinear
coordinates $(s,r,\theta )$ also as $q$ with the coordinate
indices $(1,2,3)\leftrightarrow (s,r,\theta)$, and moreover, since
it can hardly lead to a confusion we use the same notation
$\phi_{d}$ for the mappings with target spaces $\R^{3}$ and
$\Omega _{d}$ which will need later.

The geometry of $\Omega _{d}$ is naturally described in terms of
its metric tensor $(g_{ij})$; the latter is according to \cite{DE}
expressed by means of the curvature $\kappa$ and torsion $\tau$ of
$\Gamma$ in the following way
$$
g_{ij}= \left( \begin{array}{ccc}h^{2}+r^{2}\varsigma ^{2}
& 0 & r^2\varsigma  \\
0 & 1 & 0 \\ r^{2}\varsigma & 0 &  r^{2} \end{array} \right)\,,
$$
where
\begin{equation} \label{formhs}
\varsigma :=\tau -\beta_{,s}\, \quad \mathrm{and} \quad
h:=1+r\kappa \cos (\theta\!-\!\beta )\,.
\end{equation}
We use here the standard conventions
$\beta_{,s}\equiv\partial_{s}\beta$ and
$g^{ij}\equiv(g_{ij})^{-1}$. In particular, the volume element of
$\Omega_{d}$ is given by $d\Omega =g^{1/2}dq$ where $g:=\det
(g_{ij})$. The simplest situation occurs if we choose
\begin{equation} \label{Tang}
 \beta _{,s}=\tau\,,
\end{equation}
because then the tensor $g_{ij}$ takes the diagonal form $g_{ij}=
\mathrm{diag}(h^ {2}, 1,r^{2})$.
\begin{remarks} \label{prodif}
{\rm (a) It is well known that compact manifolds in $\R^{n}$ have
the tubular neighbourhood property. Thus if $\Gamma $ is a finite
$C^4$ curve then the assumption (a$\Gamma 1$) is satisfied iff
$\Gamma $ has no self-intersections. \\ [.25em]
(b) Combining the explicit formula for $g_{ij}$ with the inverse
function theorem it is easy to see that the inequality $d\|\kappa
\|_{\infty}<1$ is sufficient for $\phi_{d}$ to be locally
diffeomorphic.}
\end{remarks}

The special rotating system described above is called in the
theory of waveguides usually Tang system of coordinates. If $I$ is
finite, the functions $h,\, \dot h,\, \ddot h$ are bounded by
assumption, while in the case $I=\R$ the global boundedness has to
be assumed. The main problem, however, is that the described
construction may fail if the Frenet frame is not uniquely defined.
Hence we suppose in general that

  \begin{description}
  \item{(a$\Gamma 2$)}
for all $d>0$ small enough there is a diffeomorphism $\phi
_{d}:\:\D_{d}\to\Omega_d$ such that the corresponding metric
tensor is $g_{ij}= \mathrm{diag}(h^ {2}, 1,r^{2})$ where $h$ is
given by (\ref{formhs}) with $\beta$ which is locally bounded,
$C^{k-2}$ smooth with a possible exception of a nowhere dense
subset of $I$, and $h$ together with its first two derivatives are
bounded.
 \end{description}
While it represents a nontrivial restriction, this hypothesis can
be nevertheless satisfied for a wide class of curves without a
global Frenet frame.

\begin{example} \label{globTang}
{\rm Suppose that the curve parameter interval $I$ can be covered
by at most countable union $\bigcup_{j\in\J} I_j$ of intervals
$I_j\equiv [a_j,b_j]$ such that a pair of different $I_j,I_k$ has
in common at most one endpoint, and furthermore, either the Frenet
frame exists in $(a_j,b_j)$ or $\Gamma_{,ss}=0$ in $[a_j,b_j]$. In
the former case we assume also that limits of $n(s)$ as $s$
approaches $a_j$ and $b_j$ exist. We claim that in such a case a
diffeomorphism with a diagonal $g_{ij}$ can be constructed.

Let us describe first its building blocks. If the Frenet frame
exist in $(a_j,b_j)$ we construct a map $\phi_d^{(j)}$ on the
appropriate part of the curve by (\ref{Tangdif}) with $\beta$
replaced by a function $\beta_j$ satisfying the condition
(\ref{Tang}). On the other hand, in the straight parts we
construct $\phi_d^{(j)}$ similarly choosing a constant for
$\beta_j$ and an arbitrary pair of unit vectors forming an
orthogonal system with the tangent for $b,n$ with a proper
orientation.

The maps $\phi_d^{(j)}$ can be patched into a global
diffeomorphism by choosing properly the $\beta_j$'s. Suppose first
that the family $\{a_j\}$ of left endpoints has no accumulation
points in the interior of $I$. In that case we may identify
without loss of generality the index set $\J$ with a segment of
$\Z$, i.e. $j=-N, -N+1, \dots, M$ for some $N,M\in
\N_0\cup\{\infty\}$ and to suppose that the interval family is
ordered, $a_{n+1}=b_n$. Assume now that such a diffeomorphic map
exists on $\bigcup_{j=-n}^m I_j$. The left and right limits of the
base vector systems at the points $a_{-n}= b_{-n-1}$ and
$b_m=a_{m+1}$ exist by assumption and differ at most by a rotation
in the normal planes to $\Gamma$ at these points, so the map can
be extended to a diffeomorpism on $\bigcup_{j=-n-1}^{m+1} I_j$ by
adjusting $\beta_{-n-1}$ and $\beta_{m+1}$; notice that the
condition (\ref{Tang}) remains valid on $(a_j,b_j)$ if the
function $\beta_j$ is shifted by a constant. The sought conclusion
then follows easily by induction. On the other hand, let the set
$\{a_j\}$ have accumulation points in the interior of $I$; we call
them $\{c_k\}$ ordering them into a increasing sequence, finite or
countable. The above construction defines a diffeomorphism in each
interval $\tilde I_k:= (c_k,c_{k+1})$ between adjacent points.
Then the argument can be repeated, the role of $I_j$ being now
played by the intervals $\tilde I_k$.

Having constructed such a $\phi _{d}:\:\D_{d}\to\Omega_d$ one can
check directly whether the global boundedness conditions of the
assumption (a$\Gamma 2$) are satisfied.}
\end{example}


\subsection{Singularly perturbed Schr\"odinger operators}
\label{H}

The Hamiltonians we want to study are Schr\"odinger operators with
$s$-inde\-pen\-dent perturbations supported by the curve $\Gamma$.
Such operators can be understood as the Laplacian with specific
boundary conditions on $\Gamma$ and the aim of this section is to
make this conditions precise.

Let us assume that for a number $d$ the map $\phi_d$ satisfies
conditions (a$\Gamma 1,2$). Given $\rho\in(0,d)$ and $\theta
_{0}\in [0,2\pi)$ denote by $\Gamma_{\rho, \theta_{0}}$ the
``shifted'' curve located at the distance $\rho$ from $\Gamma$
which is defined as the $\phi_d$ image of the set
$I\times\{\rho,\theta_0\} \subset \D_d$; recall that the global
diffeomorhism $\phi_d$ exists by assumption~(a$\Gamma 2$).
Consider the Sobolev space $W^{2,2}_{\mathrm{loc}} (\Lambda
\setminus \Gamma)$, where $\Lambda$ is an open bounded or
unbounded set in $\R^3$ such that $\Omega _{d}\subseteq \Lambda$;
particularly $\Lambda$ may coincide with whole $\R^3$. Since its
elements are continuous on $\Lambda$ away of $\Gamma $, the
restriction of a function $f\in W^{2,2}_{\mathrm{loc}} (\Lambda
\setminus \Gamma)$ to the ``shifted'' curve located sufficiently
close to $\Gamma $ is well defined; we will denote it as
${f\!\upharpoonright}_{\Gamma_{\rho , \theta_{0}}}(\cdot)$. In
fact, we can regard ${f\!\upharpoonright}_{\Gamma_{\rho, \theta
_{0}}}$ as a distribution from $D^{\prime }(0,L)$ parameterized by
the distance $\rho$ and the angle $\theta_0$. We shall say that a
function $f\in W_{\mathrm{loc}}^{2,2}(\Lambda\setminus \Gamma)\cap
L^{2}(\Lambda)$ belongs to $\Upsilon _{\Omega _d}$ if the
following limits,
 \begin{eqnarray*}
 \Xi (f)(s) &\!:=\!& -\lim_{\rho \to 0}\: \frac{1}{\ln \rho }\,
 {f\!\upharpoonright}_{\Gamma_{\rho ,\theta _{0}}}(s)\,, \\
 \Omega (f)(s) &\!:=\!& \lim_{\rho \to 0}\,
 \left[{f\!\upharpoonright}_{\Gamma_{\rho ,\theta _{0}}} (s)
+\Xi (f)(s)\ln \rho \right] \,,
 \end{eqnarray*}
exist a.e. in $[0,L]$, are independent of $\theta_{0}$, and define
a pair of functions belonging to $L^{2}(0,L)$; for an infinite
curve $[0,L]$ is replaced by $\R$. We should also stress here that
the elements of $W^{2,2}_{\mathrm{loc}} (\Lambda \setminus
\Gamma)$ are in fact distributions from $D^{\prime }(\R^{3})$,
however, in the definition of $\Upsilon _{\Lambda}$ we can
naturally identify them with their canonical imbeddings into
$L^{2}(\Lambda)$.

Given a function $f\in \Upsilon_{\Lambda}$ we write $f\symbol{126}
\alpha . bc(\Gamma)$ if the limits $\Xi (f)(\cdot), \,\Omega
(f)(\cdot)$, characterizing the behavior of $f$ close to $\Gamma$
satisfy the following relation
\begin{equation} \label{boucon}
2\pi \alpha \Xi (f)(s)=\Omega (f)(s)\,.
\end{equation}
With these prerequisites we can define the singularly perturbed
Schr\"odinger operator in question through the set
$$
D(H_{\alpha ,\Gamma })=\{f\in \Upsilon
_{\R^{3}}:f\symbol{126}\alpha .bc (\Gamma )\}
$$
on which the operator $H_{\alpha ,\Gamma }:D(H_{\alpha ,\Gamma
})\rightarrow L^{2}(\R^{3})$ acts as
\begin{equation} \label{Hamilt}
H_{\alpha ,\Gamma }f(x)=-\Delta f(x)\,,\quad x\in \R^{3}\setminus
\Gamma\,.
\end{equation}
To show that $H_{\alpha ,\Gamma }$ makes sense as a quantum
mechanical Hamiltonian we will assume here that $\Gamma $ is
finite or infinite periodic. Another interesting case, that of an
infinite non-periodic curve which is asymptotically straight,
needs additional assumptions and will be discussed separately in
Sec.~\ref{infcur}.
\begin{theorem} \label{Hamsad}
Under the stated assumptions $H_{\alpha ,\Gamma }$ is
self-adjoint.
\end{theorem}
\begin{proof}
One check using integration by parts and passing to the
curvilinear system of coordinates $q=(s,r,\theta)$ in a
sufficiently small tubular neighbourhood of $\Gamma $ that the
following boundary form,
$$
\upsilon:\: \upsilon(f,g)=(H_{\alpha ,\Gamma }f,g)-(f,H_{\alpha
,\Gamma }g)
$$
vanishes for all $f,\, g\in D(H_{\alpha ,\Gamma })$, i.e. that the
operator $H_{\alpha ,\Gamma }$ is symmetric. To check its
self-adjointness we can proceed in analogy with
\cite[Thm.~4.1]{EK}. Repeating the argument presented there step
by step we derive the resolvent of $H_{\alpha ,\Gamma }$ and the
sought result then follows from \cite[Theorem~2.1]{AP}. An
alternative way is to note that $H_{\alpha ,\Gamma }$ is one of
the self-adjoint extensions discussed in \cite{Ku}. It is true
that in this paper stronger smoothness conditions for $\Gamma $
were adopted, however, the results remain valid for the $C^{4}$
class.
\end{proof}
\medskip

\noindent The operator $H_{\alpha,\Gamma}$ will be a central
object of our interest. It is natural to regard it as a
Schr\"odinger operator with the singular perturbation supported by
the curve $\Gamma$.

\begin{remarks} \label{trueX}
{\rm (a) The choice of boundary conditions (\ref{boucon}) which we
used in the construction had a natural motivation. If $\Gamma$ is
a line in $\R^3$ one can separate variables; in the cross plane we
then have the two dimensional Laplace operator with a
single-centre point interaction $-\Delta_{\alpha,\{0\}}$ which is
a well studied object -- cf.~\cite[Sec.~I.5]{AGHH}. To define it,
one considers for a function $f\in W^{2,2}_{\mathrm{loc}}
(\R^{2}\setminus \{0\}) \cap L^{2}(\R^{2})$ the following limits
$$
\tilde{\Xi}(f):=-\lim _{r\to 0}\frac{1}{\ln r }f\,, \quad
\tilde{\Omega }(f):= \lim _{r\to 0}(f+\tilde{\Xi} (f)\ln r)\,;
$$
if they are finite and satisfy the relation
\begin{equation} \label{boucoa}
2\pi \alpha \tilde {\Xi}(f)=\tilde{\Omega }(f)\,,
\end{equation}
the function $f$ belongs to the domain of
$-\Delta_{\alpha, \{0\}}$. Using the explicit form of its
resolvent it is easy to see that such an operator has for any
$\alpha\in\R$ exactly one negative eigenvalue which is given by
\begin{equation} \label{cross_ev}
\xi _{\alpha }=-4e^{2(-2\pi \alpha +\psi (1))}\,,\quad \psi
(1)=-0.577...
\end{equation}
Obviously, it coincides with the bottom of the essential spectrum
of $H_{\alpha,\Gamma}$ for a straight $\Gamma$. We know from
\cite{EK} that this property is preserved if $\Gamma$ is curved
but asymptotically straight in a suitable sense; in that case the
operator has a non-empty discrete spectrum -- cf.
Sec.~\ref{infcur}. It is also clear from the relation
(\ref{cross_ev}) and the corresponding eigenfunction
\cite[Sec.~I.5]{AGHH} that a strong
coupling corresponds to large negative values of $\alpha$. \\
[.25em]
(b) For the sake of brevity we use in analogy with (\ref{boucon})
for the boundary conditions (\ref{boucoa}) the abbreviation
$f\symbol{126}\alpha.bc (0)$, later we employ similar
self-explanatory symbols for other  conditions, Dirichlet,
Neumann, periodic, etc. }
\end{remarks}


\setcounter{equation}{0}
\section{Strong coupling asymptotics for a loop} \label{asloop}

\bigskip

In this section we will discuss in detail the strong-coupling
asymptotic behavior of the discrete spectrum in the simplest case
when $\Gamma$ is a finite closed curve satisfying the regularity
assumptions stated above; by Remark~\ref{prodif}(a) it means that
$\Gamma$ is $C^4$ and does not intersect itself.

\begin{remark} \label{}
{\rm In the following considerations we will rely on the operator
inequality $A\leq B$, where both of operators $A,B$ are
self-adjoint and bounded from below. To be precise we are going to
follow the definition from \cite[Sec.~XIII.15]{RS}, i.e. $A\leq B$
\emph{iff}
$$ q_{A}[f]\leq q_{B}[f]\,,\quad f\in Q(B)\subseteq Q(A)\,, $$
where $q_{A},q_{B}$ are the forms associated with $A,B$ having the
form domains $Q(A),Q(B)$, respectively.}
\end{remark}

Since $\Gamma$ is compact it does not influence the essential
spectrum of $H_{\alpha, \Gamma}$. This can be seen by writing
explicitly the resolvent \cite{AP} and checking that it differs
from the free one by a compact operator in analogy with the
argument used in \cite{BEKS} for $\mathrm{codim}\,\Gamma =1$.
However, there is a simpler way.

\begin{proposition} \label{essloop}
With the stated assumptions we have
$$
\sigma_{\mathrm{ess}}(H_{\alpha, \Gamma})
=\sigma_{\mathrm{ess}}(-\Delta) =[0,\infty )\,.
$$
\end{proposition}
\begin{proof}
By Neumann bracketing we can check that $\inf
\sigma_{\mathrm{ess}} (H_{\alpha, \Gamma}) =0$. Indeed, choose a
ball $\B$ such that $\Gamma$ is contained in its interior and call
$H_{\alpha ,\Gamma }^{N_{\partial \mathcal{B}}}$ the Laplace
operator in $L^{2}(\R^{3})$ with the same boundary condition on
$\Gamma $ as $H_{\alpha ,\Gamma }$ and Neumann condition at
$\partial\B$. We have $H_{\alpha, \Gamma} \ge H_{\alpha ,\Gamma
}^{N_{\partial \mathcal{B}}}$ and the spectrum of the latter is
the union of the interior and the exterior component. The first
named one is discrete and the spectrum of the other is the
non-negative halfline, so the claim follows from the minimax
principle. To show that every positive number belongs to
$\sigma(H_{\alpha, \Gamma})$ it is sufficient to construct a
suitable Weyl sequence; one can use a Weyl sequence for $-\Delta$
chosen in such a way that its elements have supports disjoint from
$\B$. \end{proof}

\medskip

Let us turn to the main subject of this section. To describe how
the discrete spectrum of $H_{\alpha }$ behaves asymptotically for
$\alpha \rightarrow -\infty $ we employ the comparison operator
defined by
\begin{equation} \label{compar}
S=-\frac{d^{2}}{ds^{2}}-\frac{\kappa (s)^{2}}{4}:\: D(S)\to
L^{2}(0,L)\,,
\end{equation}
with the domain $D(S)=\{\phi \in W^{2,2}(0,L);\phi
\symbol{126}p.bc(0,L)\}$, i.e. determined by periodic boundary
conditions, $\phi (0)=\phi (L),\phi ^{\prime }(0)=\phi ^{\prime
}(L)$. Furthermore, $ \kappa (\cdot )$ is the curvature of
$\Gamma$. It is worth to stress that $S$ acts in a different
Hilbert space than $H_{\alpha, \Gamma}$. We denote by $\mu _{j}$
the $j$-th eigenvalue of $S$. With this notations our main result
looks as follows:

\begin{theorem} \label{evloop}
(a) To any fixed $n\in\N$ there exists an $\alpha(n)\in \R$ such
that
$$
\#\sigma _{d}(H_{\alpha, \Gamma})\geq n \quad for \quad\alpha \leq
\alpha (n)\,.
$$
The $j$-th eigenvalue $\lambda _{j}(\alpha )$ of $H_{\alpha,
\Gamma}$ admits an asymptotic expansion of the following form,
$$
\lambda _{j}(\alpha )=\xi _{\alpha }+\mu _{j}+\mathcal{O}(e^{\pi
\alpha })\quad as \quad \alpha \rightarrow -\infty\,.
$$
(b) The counting function $\alpha \mapsto \,\#\sigma
_{d}(H_{\alpha, \Gamma})$ behaves asymptotically as
$$
\#\sigma _{d}(H_{\alpha })=\frac{L}{\pi }(-\xi
_{a})^{1/2}(1+\mathcal{O} (e^{\pi \alpha })).
$$\
\end{theorem}

\bigskip

\noindent The proof of the theorem is divided into several steps
which we will describe subsequently in the following sections. It
is also worth to stress here that the error term
$\mathcal{O}(e^{\pi \alpha })$ is not uniform with respect to $j$;
this will be clear from Lemma~\ref{eivalj} below.

\subsection{Dirichlet-Neumann bracketing}

Our aim is to estimate the operator $H_{\alpha, \Gamma}$ in the
negative part of its spectrum from both sides by means of suitable
operators acting in a tubular neighbourhood $\Omega_{d}$ of
$\Gamma$, with $d$ sufficiently small to make the assumptions
(a$\Gamma 1,2$) satisfied. The first step in obtaining the
estimating operators is to impose additional Dirichlet and Neumann
condition at the boundary of $\Omega_{d}$. Let thus the operators
$H_{\alpha, \Gamma}^j,\: j=D,N$, in $L^{2}(\Omega_{d})$ act as the
Laplacian with the domains given respectively by $D(H_{\alpha,
\Gamma}^j)=\{f\in \Upsilon _{\Omega _{d}}:\: f\symbol{126}\alpha
.bc (\Gamma ),\: f\symbol{126}j.bc(\partial \Omega _{d})\}$; it is
straightforward to check that operators $H_{\alpha, \Gamma}^j$ are
self-adjoint. Now the well-known result \cite[Sec.~XIII.15]{RS}
says that
$$
-\Delta _{\Sigma _{d}}^{N}\oplus H_{\alpha, \Gamma}^{N}\leq
H_{\alpha, \Gamma}\leq -\Delta _{\Sigma _{d}}^{D}\oplus H_{\alpha,
\Gamma}^{D}\,,\quad \Sigma _{d}:= \R^{3} \setminus
\overline{\Omega }_{d}\,.
$$
What is important is that the operators $-\Delta _{\Sigma _{d}}^j$
corresponding to the exterior of $\Omega_d$ do not contribute to
the negative part of the spectrum because they are both positive
by definition.

It is convenient to express the operators $H_{\alpha, \Gamma}^j$
in the curvilinear coordinates $q=(s,r,\theta );$ this can be done
by means of the unitary transformation
$$
Uf=f\circ \phi _{d}:L^{2}(\Omega _{d})\to
L^{2}(\D_{d},g^{1/2}dq)\,,\quad \D_{d}=[0,L]\times B_{d}\,;
$$
recall that the global diffeomorhism $\phi_d$ exists by
assumption~(a$\Gamma 2$). Then the operators $\tilde H_{\alpha,
\Gamma}^j:= UH_{\alpha, \Gamma}^j U^{-1}$ act as
$$
f(x) \mapsto -(g^{-1/2}\partial _{i}g^{1/2}g^{ij}\partial
_{j}f)(x) \quad \mathrm{for} \quad x\in \Omega_{d}\setminus \Gamma
$$
with the domains $\{f\in \Upsilon _{\Omega _{d}}:\:
f\symbol{126}\alpha .bc( \Gamma ),f\symbol{126}j.bc( \omega_{r}(d)
),\: f\symbol{126}p.bc(\omega_{s}(0),\omega_{s}(L))\}$,
respectively, where we have introduced the notation
$$
\omega_{q_i}(t):=\{q\in \overline{\mathcal{D}}_{d}:q_{i}=t\}.
$$
To simplify it further we remove the weight $g^{1/2}$ appearing in
the inner product of the space $ L^{2}(\D_{d},g^{1/2}dq)$. This is
done by means of the another unitary map,
$$
\hat{U}:L^{2}(\D_{d},g^{1/2}dq)\to L^{2}(\D_{d},dq)\,,\quad
\hat{U}f:=g^{1/4}f\,;
$$
the images of $\tilde H_{\alpha, \Gamma}^j$ will be denoted as
$\hat H_{\alpha, \Gamma}^j= \hat{U}\tilde H_{\alpha, \Gamma}^j
\hat{U} ^{-1}$. The aim of these unitary transformations is to
find a representation where the eigenvalues -- which we need to
estimate the eigenvalues of $H_{\alpha, \Gamma}$ by means of the
minimax principle -- are easy to analyze. A straightforward
calculation analogous to that performed in \cite{DE} yields
explicit formulae for $\hat H_{\alpha, \Gamma}^j,\: j=D,N$, which
both act as\footnote{We employ the usual convention that summation
is performed over repeated indices keeping in mind that $(g^{ij})$
is diagonal.}
$$
-\partial _{i}g^{ij}\partial _{j}-\frac{1}{4}r^{-2}+V\,,
$$
where $V$ is the effective potential given by
\begin{equation} \label{unihat}
V=g^{-1/4}(\partial _{i}g^{ij}(\partial _{j}g^{1/4}))
+\frac{1}{4}r^{-2},
\end{equation}
while their domains are different,
\begin{eqnarray*}
D(\hat H_{\alpha, \Gamma}^D) &\!=\!& \{f\in\Upsilon
_{\mathcal{D}_{d}}:\: g^{-1/4}f\symbol{126} \alpha .bc(\Gamma ),\:
f\symbol{126}p.bc(\omega_{s}(0),\omega_{s}(L))\,,
\\ && \phantom{i}
f\symbol{126}D.bc( \omega_{r}(d) )\}\,, \\
D(\hat H_{\alpha, \Gamma}^N) &\!=\!& \{f\in\Upsilon
_{\mathcal{D}_{d}}:\: g^{-1/4}f\symbol{126}\alpha .bc(\Gamma ),\:
f\symbol{126}p.bc(\omega_{s}(0),\omega_{s}(L))\,, \\
&& \phantom{i} (\partial _{r}f)_{r=d}
=-[(g^{1/4}\partial_{r}g^{-1/4})f]_{r=d}\}\,,
\end{eqnarray*}

\begin{remark} \label{alphaG}
{\rm Notice that the boundary conditions satisfied by functions
from $D(\hat H_{\alpha, \Gamma}^j)$ on the curve $\Gamma$ can be
written in a simpler way. Since only the leading term in
$g^{-1/4}$ is important as $r\to 0$, they are equivalent to
$r^{-1/2} f\symbol{126}\alpha .bc(\Gamma )$. Notice also that
while the Dirichlet boundary condition at $\partial \Omega_d$
persists at the unitary transformation, the Neumann one is changed
by $\hat U$ into a mixed boundary condition. }
\end{remark}

\subsection{Estimates by operators with separated
variables }

\bigskip

While the operators $\hat H_{\alpha, \Gamma}^j,\, j=D,N$, give the
two-sided bounds for the negative eigenvalues of $H_{\alpha }$,
they are not easy to handle. This is why we pass to a cruder, but
still sufficient estimate by operators with separated variables.

In the first step we will make the boundary conditions in the
lower bound independent of the coordinates. The boundary term
involved in the definition of $D(\hat H_{\alpha, \Gamma}^N)$
depends on $s$ and $\theta$. We replace the corresponding
coefficient by $M:= \left\| g^{1/4}\partial _{r}g^{-1/4}
\right\|_{L^{\infty} (\omega_r(d))}$ passing thus to the operator
$$
\dot H_{\alpha, \Gamma}^{-}:=-\Delta_{h}\otimes I +I \otimes
(-\Delta _{\alpha}^{-}) +V\leq \hat H_{\alpha, \Gamma}^N\,
$$
on $L^{2}(0,L) \otimes L^{2}(B_{d})$, where $-\Delta
_{h}:=-\partial _{s}h^{-2}\partial _{s}:\: D(S)\to L^{2}(0,L)$ and
 \begin{eqnarray*}
-\Delta _{\alpha }^{-} &\!:=\!& -\partial _{r}^{2}-r^{-2}\partial
_{\theta}^{2} -\frac{1}{4}r^{-2}:\: D(\Delta _{\alpha }^{-}) \to
L^{2}(B_{d})\,, \\
D(\Delta _{\alpha }^{-}) &\!:=\!& \{f\in W^ {2,2}_{\rm{loc}}(B_{d}
\setminus \{0\}):\: \Delta _{\alpha }^{-}f \in L^2(B_d),\:
r^{-1/2}f\symbol {126}\alpha .bc(0),\\ && (\partial _{r} f)|_{r=d}
=Mf|_{r=d}\}
 \end{eqnarray*}
with the boundary condition at the centre of the circle written in
the simplified form mentioned in Remark~\ref{alphaG}. The upper
bound contains no boundary term depending on $s$ or $\theta$ so we
can put
$$
\dot H_{\alpha, \Gamma}^{+} = \hat H_{\alpha, \Gamma}^D =
-\Delta_{h}\otimes I +I \otimes (-\Delta _{\alpha}^{+}) +V
$$
which acts in the same way but the above mixed boundary condition
on $\partial B_d$ is replaced by the Dirichlet condition.

The next estimate concerns the effective potential $V$ given by
(\ref{unihat}); by a straightforward calculation \cite{DE} we can
express it in terms of the curvature together with the function
$h$ and its two first derivatives with respect to the variable $s$
as follows,
\begin{equation}  \label{forexV}
V=-\frac{\kappa ^{2}}{4h^{2}}+ \frac{h_{,ss}}{2h^{3}} -\frac{
5(h_{,s})^{2}}{4h^{4}}\,.
\end{equation}
It is important that up to an $\mathcal{O}(d)$ term this
expression coincides with the potential involved in the comparison
operator $S$. Indeed, since $h$ is continuous on a compact set and
thus bounded, by (\ref{formhs}) there exists a positive $C_{h}$
such that the inequalities
$$
C_{h}^{-}(d)\leq h^{-2}\leq C_{h}^{+}(d)\quad \mathrm{with} \quad
C_{h}^{\pm }(d):= 1\pm C_{h}d,
$$
hold for all $d$ small enough. Since $\Gamma$ is $C^4$ by
assumption, the derivatives $h_{,s}$ and $h_{,ss}$ are also
bounded; hence (\ref{forexV}) yields the estimate
$$
\left| V+\frac{\kappa ^{2}}{4}\right| \leq C_{V}d
$$
with a positive $C_{V}$ valid on $\D_d$ for all sufficiently small
$d$. At the same time, we can apply the above bounds for $h^{-2}$
to the longitudinal part of the kinetic term. Putting all this
together we get
$$
L_{d}^{-}\otimes I \leq -\Delta _{h}\otimes I +V \leq
L_{d}^{+}\otimes I\,,
$$
where
$$
L_{d}^{\pm }:= -C_{h}^{\pm }\frac{d^{2}}{ds^{2}}-\frac{\kappa
^{2}}{4 }\pm C_{V}d:\: D(S)\to L^{2}(0,L)\,.
$$
Summarizing the above discussion, we can introduce a pair of
operators with the longitudinal and transverse components
separated, namely
\begin{equation} \label{sepdec}
B^{\pm}_{\alpha }:=L_{d}^{\pm }\otimes I +I\otimes (-\Delta
_{\alpha }^{\pm }) \quad \mathrm{on} \quad L^{2}(0,L) \otimes
L^{2}(B_{d})\,,
\end{equation}
which give the sought two-sided bounds, $\pm \dot H_{\alpha,
\Gamma}^{\pm} \le \pm B^{\pm}_{\alpha }$.

\subsection{Component eigenvalues estimates}
\label{lontra}

In the next step we have to estimate the eigenvalues of
$L_{d}^{\pm}$ and $-\Delta ^{\pm}_{\alpha }$. Let us start with
the longitudinal part. It is easy to check the identity
$$
L_{d}^{\pm}= C_{h}^{\pm}(d) S \pm \left(C_{V} +C_h {\kappa^2\over
4} \right)d\,;
$$
combining it with the minimax principle and the fact that the
eigenvalues of $S$ behave as $({2\pi\over L})^2 \ell^2 +
\mathcal{O}(1)$ as $\ell\to\pm\infty$, we arrive at the following
conclusion:
\begin{lemma} \label{eivalj}
There is a positive $C$ such that the eigenvalues $l^{\pm}_{j}(d)$
of $L^{\pm}_d$, numbered in the ascending order, satisfy the
inequalities
\begin{equation} \label{asylpm}
|l^ {\pm}_{j}(d)-\mu _{j}|\leq Cj^ {2}d
\end{equation}
for all $j\in\N$ and $d$ small enough.
\end{lemma}

The transverse part is a bit more involved. Our aim is to show
that in the strong-coupling case the influence of the boundary
conditions is weak, i.e. that the negative eigenvalues of the
operators $-\Delta_{\alpha }^{\pm}$ do not differ much from the
number (\ref{cross_ev}).

\begin{lemma} \label{eivt+-}
There exist positive numbers $C_{i}$, $1\leq i\leq 4$, such that
each one of the operators $-\Delta _{\alpha }^ {\pm}$ has exactly
one negative eigenvalue $t_{\alpha }^{\pm}$ which satisfies
\begin{equation} \label{eigvat}
\xi_{\alpha }-S(\alpha ) < t_{\alpha }^{-}<\xi _{\alpha }<
t_{\alpha } ^ {+}< \xi _{\alpha }+S(\alpha )
\end{equation}
for $\alpha$ large enough negative, where
$$
S(\alpha ):=C_{1}\zeta _{\alpha }^{2}\sqrt {d\zeta _{\alpha }}\exp
(-C_{2}d\zeta _{\alpha })
$$
with $\zeta _{\alpha }:=(-\xi_{\alpha })^{1/2}$, provided
$\,d\zeta _{\alpha }>C_{3}$ and $\,dM<C_{4}$.
\end{lemma}
\begin{proof}
Let us start with the eigenvalue of the operator $-\Delta ^
{+}_{\alpha }$ involved in the upper bound; the argument will be
divided into four parts.

\emph{1.~step:} We will show that the number $-k^ {2}_{\alpha }$
with $k_{\alpha }>0$ is an eigenvalue of $-\Delta ^ {+}_{\alpha }$
\emph{iff} $\,k_{\alpha }$ is a solution of the equation
\begin{equation} \label{eigequ}
x=\zeta _{\alpha }\eta(x)\,,
\end{equation}
where  $\zeta _{\alpha }$ has been defined above and $\eta $ is
the function given by
\begin{equation} \label{funeta}
\eta:\:\R _{+}\to \R_{+}\,,\quad \eta(x)=\exp \left(-\frac{
K_{0}(xd)}{I_{0}(xd)}\right)\,;
\end{equation}
the symbols $K_{0},\,I_{0}$ denote the Macdonald and the modified
Bessel function, respectively \cite{AS}. To verify this claim we
note that the eigenfunction $\varphi$ of $-\Delta _{\alpha }^{+}$
corresponding to $-k_{\alpha }^{2}$ is a linear combination
$$
\varphi (r)=D_{1}I_{0}(k_{\alpha } r)r^{1/2}+D_{2}K_{0}(k_{\alpha}
r) r^{1/2}
$$
with the coefficients $D_{1},\,D_{2}$ chosen in such a way that
the conditions following from $\varphi \symbol{126} D.bc(\partial
B_{d})$ and $r^{-1/2}\varphi \symbol {126} \alpha .bc(0)$ are
satisfied. Using the behavior of $K_{0},\,I_{0}$ at the origin
\begin{equation} \label{beK0H0}
K_{0}(\rho)= -\ln \frac{\rho}{2}+\psi(1)+\mathcal{O}(\rho ) \quad
\mathrm{and} \quad I_{0}(\rho ) =1+\mathcal{O}(\rho )\,,
\end{equation}
as $\rho\to 0$, we can readily check that $\varphi$ fulfils the
needed boundary conditions \emph{iff} $\,(D_{1},D_{2})\in \ker
M(\alpha )$, where $M(\alpha )$ is the matrix given by
$$
M_{ij}(\alpha )= \left( \begin{array}{cc}I_{0}( k_{\alpha }d)
& K_{0}(k_{\alpha }d)  \\
1 & \omega (\alpha ,k_{\alpha })
\end{array} \right)
$$
with $\omega (\alpha ,k_{\alpha }):= \psi (1)-2\pi \alpha -\ln
(k_{\alpha }/2) $. Of course, the condition $\mathrm{ker}M(\alpha
)\neq \emptyset$ is equivalent to $\det M(\alpha )=0$; the latter
holds \emph{iff} $k_{\alpha }$ is a solution of (\ref{eigequ}).

\emph{2.~step:} Our next aim is show that the equation
(\ref{eigequ}) has at least one solution for $-\alpha $
sufficiently large, and moreover, that such a solution $k_{\alpha
}$ satisfies the inequalities
\begin{equation} \label{inekal}
\tilde{C}\zeta _{\alpha } <k_{\alpha }< \zeta _{\alpha }
\end{equation}
with $\tilde{C}\in (0,1)$ independent of $\alpha$. Using again
(\ref{beK0H0}) together with the asymptotic behavior of the
functions $K_{0},I_{0}$ at infinity, we get for a fixed $\alpha $
$$
\zeta _{\alpha }\eta(x)\rightarrow \zeta _{\alpha } \quad
\mathrm{as} \quad x\rightarrow \infty
$$
and
\begin{equation} \label{asy}
\zeta _{\alpha }\eta(x)= g_{\alpha ,d}x+\mathcal{O} (x^{2})\quad
\mathrm{as} \quad x\rightarrow 0\,,
\end{equation}
where $g_{\alpha ,d}:= \frac{1}{2}\,\e^{-\psi (1)}d \zeta_{\alpha
}$. It is clear that the error term is uniform with respect to
$\alpha $ over finite intervals only, however, if
\begin{equation} \label{auzeta}
g_{\alpha ,d}>1
\end{equation}
then the equation (\ref{eigequ}) has obviously at least one
solution. The second inequality in (\ref{inekal}) holds trivially
because $\eta(x)<1$ for any $x>0$. Let us assume that the first
one is violated. This means that there is a sequence
$\{\alpha_n\}$ with $\alpha _{n}\to -\infty $ as $n\to \infty $
such that $\eta(k_{\alpha _{n}})\to 0$ as $ n\to \infty $. This
may happen only if the $k_{\alpha _{n}}$ tends to the singularity
of $K_{0}$, in other words if $k_{\alpha _{n}}\to 0 $ holds as
$n\to \infty $. However, the inequality (\ref{auzeta}) is valid
for $\alpha_n$ with $n$ large enough, thus small $k_{\alpha _{n}}$
can not in view of the asymptotics (\ref{asy}) be a solution of
(\ref{eigequ}) in contradiction with the assumption.

\emph{3.~step:} To show that there exists only one solution of
(\ref{eigequ}) it suffices to check that the function $h_{\alpha }
:\R_{+}\mapsto \R$,
$$
h_{\alpha }(x)=x-\zeta_{\alpha }\eta(x)\,,
$$
is strictly monotonous for $x\in (\tilde{C}\zeta _{\alpha },\zeta
_{\alpha })$ and $-\alpha $ sufficiently large. Using again the
behavior of $K_{0},\,I_{0}$ at large values of the argument we
find that the derivative $\eta ^{\prime }(x)\to 0 $ as $x\to
\infty$ which implies the result.

\emph{4.~step:} It remains to show that the eigenvalue $t_{\alpha
}^{+} = -k_{\alpha }^{2}$ satisfies the second one of the
inequalities
\begin{equation} \label{+Salph}
\xi_{\alpha }<-k_{\alpha }^{2}<\xi_{\alpha }+S(\alpha )\,.
\end{equation}
Since the functions $-K_{0},\, I_{0}$ are increasing and
$I_0(0)=1$ we get from (\ref{inekal}) the estimate
$$
\eta(k_{\alpha })\geq \exp \left(-K_{0}(\tilde{C}\zeta_{\alpha }
d)\right).
$$
Putting now $\tilde {S}(\alpha )=\left(1-\exp
\left(-2K_{0}(\tilde{C}\zeta_{\alpha } d)\right)\right)\zeta
_{\alpha }^{2}$ and using the asymptotic behavior of $K_{0}$ at
large distances one finds that
$$
\tilde{S}(\alpha )\leq \tilde{C}_{1}\zeta _{\alpha }^{2}\sqrt
{d\zeta _{\alpha }}\exp (-\tilde{C}_{2}d\zeta _{\alpha }) \quad
\mathrm{as} \quad \alpha \to -\infty
$$
holds with suitable constants $\tilde{C}_{1},\,\tilde{C}_{2}$ and
the inequality (\ref{+Salph}) is satisfied which concludes the
proof for the operator $-\Delta _{\alpha }^{+} $.

Let us turn to the operator $-\Delta ^{-}_{\alpha }$. The argument
is similar, so we just sketch it with the emphasis on the
differences. The number $t^{-}_{\alpha }=-k_{\alpha }^{2}$ is an
eigenvalue of $-\Delta ^{-}_{\alpha }$ \emph{iff} $k_{\alpha }$ is
a solution of the equation
\begin{equation} \label{eigeq2}
x=\zeta _{\alpha }\tilde{\eta }(x)\,,
\end{equation}
where $\tilde{\eta }:\R _{+}\to \R_{+}$ is the function given by
$$ 
\tilde{\eta }(x)=\exp \left(-\frac{
S_{K}(xd)}{S_{I}(xd)}\right)\,,\quad S_{F}(xd)=\tilde{F}_{1}
(xd)xd+w_{d}F_{0}(xd)
$$ 
for $F=K,I$, where $\tilde{I}_{1}=I_{1}$, $\tilde{K}_{1}=-K_{1}$
and $w_{d}:= \frac{1}{2}-Md$; we assume that
\begin{equation} \label{condwd}
w_{d}>0\,.
\end{equation}
To proceed further, we employ again the asymptotics of functions
$I_{n},K_{n}$, $n=0,1$, for $x\to 0$ and at large values of the
argument. It is easy to see that the behavior of $x\mapsto
\frac{S_{K}(xd)}{S_{I}(xd)}$ for small $x$ is dominated by that of
$K_{0}(\cdot)$. Thus mimicking the second step of the above
argument we can show that the equation (\ref{eigeq2}) has at least
one solution for $-\alpha $ sufficiently large provided that
assumption (\ref{condwd}) is satisfied. Repeating the third step
we can check that the solution $k_{\alpha }$ is unique for
$-\alpha $ sufficiently large. By \emph{reduction ad absurdum}, as
in the second step, we can also prove that there exists $\hat{C}$
such that $\hat{C}\zeta _{\alpha }<k_{\alpha }$, which means that
$k_{\alpha }\to \infty $ as $\alpha \to -\infty $. The constant
can be made more specific: using the fact that the term
$-dx\,K_{1}(dx)$ dominates the behavior of $S_{K}(dx)$ for large
$x$ and $S_{I}>0$ we get $\tilde{\eta} >1$ for $-\alpha $
sufficiently large, i.e.
$$
\zeta_{\alpha }<k_{\alpha }\,.
$$
Using properties of the special functions involved here we also
find that
$$
\tilde{\eta }(k_{\alpha })\leq \exp(\dot{C}K_{1}(d\zeta_{\alpha
})d\zeta _{\alpha })
$$
holds for any $\dot{C}$ satisfying $(w_{d})^{-1}+ (d\zeta _{\alpha
})^{-1}<\dot{C}$. Thus proceeding similarly as in the fourth step
we infer that there are constants $\breve{C}_{1}, \,\breve{C}_{2}$
such that
\begin{equation} \label{-Saplh}
\xi_{\alpha }-\tilde {S}(\alpha ) < -k_{\alpha }^{2}<\xi _{\alpha
}\,,
\end{equation}
where
$$
\tilde{S}(\alpha )\leq \breve{C}_{1}\zeta _{\alpha }^{2}\sqrt
{d\zeta _{\alpha }}\exp (-\breve{C}_{2}d\zeta _{\alpha }) \quad
\mathrm{as} \quad \alpha \to -\infty \,.
$$
Finally putting together (\ref{auzeta}), (\ref{+Salph}),
 (\ref{condwd}) and (\ref{-Saplh}) we get the claim with
$C_{1}:=\max\{\tilde{C}_{1},\breve{C_{1}}\}$ and
$C_{2}:=\min\{\tilde{C}_{2},\breve{C_{2}}\}$.
\end{proof}


\subsection{ Proof of Theorem~\ref{evloop} for a loop}

Suppose now that $\Gamma$ is a closed curve. The result will
follow from combination of the above estimates. We have to couple
the width of the neighbourhood $\Omega _ {d}$ and the coupling
constant $\alpha $ in such a way that $d$ shrinks properly to zero
as $\alpha\to-\infty$. This is achieved, e.g., by choosing
\begin{equation} \label{asyalp}
d(\alpha )=\e ^{\pi \alpha }\,.
\end{equation}
\emph{Proof of (a):} To find the asymptotic behavior of
eigenvalues $\lambda _{j}(\alpha )$ of $H_{\alpha ,\Gamma }$ we
will rely on the decomposition (\ref{sepdec}), according to which
we know that the negative eigenvalues of $H_{\alpha ,\Gamma }$ are
squeezed between $l_{j}^{\pm}(d)+t_{\alpha } ^{\pm}$. Since the
operators $-\Delta ^{\pm}_{\alpha }$ have a single negative
eigenvalue, the sought values $\lambda _{j}(\alpha )$ are ordered
in the same way as $l^{\pm}_{j}(d)$ are. Combining (\ref{asyalp})
with the results of Lemmas~\ref{eivalj} and \ref{eivt+-} we get
for the upper and lower bound
$$
l^{\pm}_{j}(d(\alpha ))+t_{\alpha }^{\pm}=\xi_{\alpha }+\mu_{j}+
\mathcal{O}(e^{\pi\alpha })\quad \mathrm{as}\quad \alpha \to
-\infty \,,
$$
and of course, the same asymptotics holds for $\lambda _{j}(\alpha
)$. Clearly, to a given integer $n$ there exists $\alpha (n)\in
\R$ such that $l^{+}_{n}(d(\alpha ))+t^{+}_{\alpha }<0$ is true
for all
$\alpha \leq \alpha (n)$; this completes the proof of (a). \\
[.5em]
\emph{Proof of (b):} Using the above asymptotic estimates and
Lemma~\ref{eivalj} we get
\begin{equation} \label{estimN}
\nu^{-}_{j}(\alpha )\leq \lambda _{j}(\alpha )\leq
\nu^{+}_{j}(\alpha )\,,
\end{equation}
where
$$
\nu^{\pm }_{j}(\alpha ):=\xi_{\alpha }+j^{2}\left
(\left(\frac{2\pi } {L}\right)^{2}+ \mathcal {O}(\e^{\pi \alpha
})\right)\pm v
$$
and $ v=4^{-1}\|\kappa ^{2}\|_{\infty }$. Combining this with the
minimax principle we arrive at the two-sided estimate
$$
\sharp \{ j\in \Z:\:\nu^{+}_{j}(\alpha )<0\} \leq \sharp \sigma
_{d}(H_{\alpha })\leq \sharp \{ j\in \Z :\:\nu^{-}_{j}(\alpha
)<0\}\,,
$$
which implies
$$
\sharp \sigma _{d}(H_{\alpha })=\frac{L}{\pi }(-\xi _{\alpha})^{1/2}(1+\mathcal{O}%
(\e^{\pi \alpha }))\,.
$$\

\subsection{A curve with free ends} \label{freeend}

The part (b) of Theorem~\ref{evloop} does not require $\Gamma$ to
be a closed curve. One can repeat the argument with a small
modification taking for $\Omega_d$ a closed tube around $\Gamma $
bordered by the additional ``lid'' surfaces normal to $\Gamma $ at
its ends. Thus instead of $S$ we have a pair of comparison
operators $S^{i}=-\frac{d^{2}}{ds^{2}}-\frac{\kappa ^{2}}{4}$ on
$L^{2}(0,L)$ with $D(S^{i}):=\{f\in W^{2,2}(0,L),
f\symbol{126}i.bc\}$, $i=D,N$, which give in the same way as above
estimates for the eigenvalues $\lambda _{j}(\alpha )$ of
$H_{\alpha ,\Gamma }$ as $\alpha \to -\infty $, namely
$$
\xi_{\alpha }+\mu^{N}_{j}+\mathcal{O}(\e^{\pi \alpha })\leq
\lambda _{j}(\alpha )\leq \xi_{\alpha
}+\mu^{D}_{j}+\mathcal{O}(\e^{\pi \alpha }),
$$
where $\mu ^{i}_{j},\: j=1,2,\dots\,$, denote the eigenvalues of
$S^{i}$. The fact that the latter are different for the Dirichlet
and Neumann condition does not allow us to squeeze $\lambda
_{j}(\alpha )$ sufficiently well to get its asymptotics in analogy
with the claim (a) of the theorem. On the other hand, the behavior
of $\mu^{D}_{j}-\mu^{N}_{j}$ as $j\to\infty$ allows us to find an
asymptotic estimate for the counting function. Recall that the
eigenvalues of $-\Delta ^{i}=-\frac {d^{2}}{ds^{2}}: D(S^{i})\to
L^{2}(0,L)$ are of the form $s_{j}^{i}= j^{2}(\frac{\pi}{L})^{2}$,
where $j\in \N$ for $i=D$ and $j\in \N\cup \{0\}$ for $i=N$; thus
in analogy with (\ref{estimN}) we can define the functions
$$
\nu^{\pm }_{j}(\alpha ):=\xi_{\alpha
}+j^{2}_{\pm}\left(\left(\frac{\pi } {L}\right)^{2}+ \mathcal
{O}(\e^{\pi \alpha })\right)\pm v \,,
$$
where $j^{+}=j$ and $j^{-}=j-1$ with $j\in \N $, which give a
two-sided bound for $\lambda _{j}(\alpha )$. Combining it with the
minimax principle we arrive again at the formula
$$
\sharp \sigma _{d}(H_{\alpha, \Gamma})=\frac{L}{\pi }(-\xi
_{\alpha})^{1/2}(1+\mathcal{O} (\e^{\pi \alpha })) \quad
\mathrm{as} \quad \alpha\to -\infty\,.
$$
\vspace{.5em}
\begin{remark} \label{finloop}
{\rm While Theorem~\ref{evloop} was formulated for a single finite
curve, which may not be closed for part (b), the argument easily
extends to any $\Gamma$ which decomposes into a finite disjoint
union of such curves, up to the eigenvalue numbering. The latter
may be ambiguous in case that the corresponding operator $S$,
which is now an orthogonal sum of components of the type
(\ref{compar}), exhibits an accidental degeneracy in its spectrum.
}
\end{remark}


\setcounter{equation}{0}
\section{Infinite asymptotically straight curves} \label{infcur}

We know from \cite{EK} that the operator $H_{\alpha, \Gamma}$ has
a nonempty discrete spectrum if $\Gamma$ is an infinite $C^4$
curve which is non straight but it is asymptotically straight in
the following sense
 \begin{description}
 \item{(a$\Gamma _\mathrm{inf}$1)} for all $s\in \R$ we have
 $|\kappa (s)|\leq M|s|^{-\beta }$,
 where $\beta >5/4$ and $M>0$.
  \end{description}
Moreover one has to assume that
 \begin{description}
 \item{(a$\Gamma _\mathrm{inf}$2)} there exists a constant $c\in (0,1)$
 such that $|\gamma(s)-\gamma(s^{\prime })|\geq c|s-s^{\prime }|$.
  \end{description}
If these conditions are satisfied then the operator $H_{\alpha
,\Gamma }$ is self-adjoint and
$$
\sigma _{\mathrm{ess}}(H_{\alpha ,\Gamma })=[\xi_{\alpha },\infty
)\,, \quad \sigma _{\mathrm{d}}(H_{\alpha ,\Gamma })\neq
\emptyset\, .
$$
Since the infinite curve has no free ends, the asymptotics of
eigenvalues of $H_{\alpha ,\Gamma }$ for $\alpha \to -\infty $ can
be found in the same way as for the loop. We employ the comparison
operator which now takes the form
$$
S=-\frac{d^{2}}{ds^{2}}-\frac{1}{4}\kappa (s)^{2}:\: D(S)\to
L^{2}(\R)
$$
with the domain $D(S)$ equal to $W^{2,2}(\R)$. It is a
Schr\"odinger operator on line with a potential which is purely
attractive provided $\kappa\ne 0$, and therefore
$$
\sigma _{\mathrm{d}}(S)\neq \emptyset \,.
$$
On the other hand, in view of the assumed decay of curvature as
$|s|\to\infty$ the number $N:=\sharp \sigma _{\mathrm{d}}(S)$ is
finite \cite[Thm.~XIII.9]{RS}. Using the symbol $\mu _{j}$ for the
$j$-th eigenvalue of the operator $S$ we get the following result.
\begin{theorem}
Under the above stated assumptions there is $\alpha _{0}\in\R$
such that $\sharp \sigma _{\mathrm{d}}(H_{\alpha ,\Gamma })=N$
holds for all $\alpha <\alpha _{0}$. Moreover, the $j$-th
eigenvalue $\lambda_{j}(\alpha )$ of $H_{\alpha ,\Gamma }$,
$j=1,\dots,N$, admits the asymptotic expansion
$$
\lambda_{j}(\alpha )=\xi_{\alpha
}+\mu_{j}+\mathcal{O}(\e^{\pi\alpha}) \quad \mathrm{as} \quad
\alpha\to -\infty\,.
$$
\end{theorem}
\vspace{.5em}

\noindent Since the proof fully analogous to that of Theorem
\ref{evloop} we omit details.


\setcounter{equation}{0}
\section{Spectrum for an infinite periodic curve} \label{asperiod}

\subsection{The Floquet--Bloch decomposition  }

Now we turn our attention to Hamiltonians with singular
perturbations supported by a periodic $C^{4}$ curve without
self-intersections. In other words we assume that there is a
vector $\mathbf{K}_1 \equiv \mathbf{K}\in\R^{3}$ and a number
$L>0$ such that
$$
\gamma (s+L)=\mathbf{K}+\gamma (s)\quad \mathrm{for\;all}\quad
s\in\R\,.
$$
Of course, we can always choose the Cartesian system of
coordinates such that $\mathbf{K} =(K,0,0)$ with $K>0$, and
$\gamma (0)=0$. As usual in periodic situations we decompose the
space $\R^{3}$ according to the periodicity of $\Gamma $. To this
aim we define the basic period cell as
\begin{equation} \label{percell}
\mathcal{C}_{0}\equiv\mathcal{C}:=\{x:x=\sum
_{i=1}^{3}t_{i}\mathbf{K}_{i},\; t_{1}\in [0,1),\, t_{i}\in \R,\,
i=2,3\, \}\,,
\end{equation}
where $\{\mathbf{K}_{i}\}_{i=1}^{3}$ are linearly independent
vectors in $\R^{3}$; without loss of generality we may suppose
that $\mathbf{K}_{2} \perp \mathbf{K}_{3}$. Then the translated
cells $\mathcal{C}_{n}:=\mathcal{C}+n\mathbf{K}$, where $n\in \Z$,
are mutually disjoint for different values of the index and
$\R^{3}=\bigcup _{n\in \Z}\mathcal{C}_{n}$. As in the previous
section we assume that $\Gamma $ has no self-intersections.
However, to proceed further we need an additional assumption,
namely
 \begin{description}
 \item{(a$\Gamma _\mathrm{per}$)} $\:$ the restriction of $\Gamma
 _{\mathcal{C}}:= \mathcal{C}\cap \Gamma $ to the interior of
 $\mathcal{C}$ is connected.
  \end{description}
Let us note the choice of the point $s=0$ is important in checking
the assumption (a$\Gamma _\mathrm{per}$), and for the same reason
we do not require generally that $\mathbf{K}_{1} \perp
\{\mathbf{K}_{2},\mathbf{K}_{3}\}$ (see also Remark~\ref{crochet}
below).

While a smooth periodic curve without self-intersections satisfies
(a$\Gamma 1$), the property (a$\Gamma _\mathrm{per}$) ensures that
we can choose a neighbourhood of $\Gamma_{\mathcal{C}}$ which is
connected set contained in $\mathcal{C}$; this is important for
the construction described below. In view of Theorem~\ref{Hamsad}
the Hamiltonian with the singular perturbation supported by
$\Gamma $ is well defined as a self-adjoint operator in $
L^{2}(\R^{3})$. To perform the Floquet--Bloch reduction for
$H_{\alpha ,\Gamma }$ we decompose first the state Hilbert space
into a direct integral
$$
\mathcal{H}=\int _{[-\pi/K,\pi/K )}^{\oplus }\mathcal{H}^{\prime
}\, d\theta \,, \quad \mathcal{H}^{\prime }:=L^{2}(\mathcal{C})\,.
$$
It is a standard matter to check that the operator $U:\,L^{2}
(\R^{3})\to \mathcal{H}$ given by
\begin{equation} \label{operaU}
(Uf)_{\theta }(x)=\frac{1}{(2\pi)^{1/2}}\sum _{n\in \Z}\e
^{-i\theta Kn}f(x+n\mathbf{K})
\end{equation}
on $f\in C^{\infty }_{0}(\R^{3})$ acts isometrically, so it can be
uniquely extended to a unitary operator on the whole $L^{2}
(\R^{3})$. We will say that the function $f\in
C^{2}(\mathcal{C}\backslash \Gamma _{\mathcal{C}})$ \emph{belongs
to} $\Upsilon_{\alpha }(\theta )$ if it satisfies the condition
$$f\symbol{126}\alpha .bc(\Gamma _{\mathcal{C}})\,,$$
and furthermore, for all $x$ such that both $x$ and $x+\mathbf{K}$
belong to $\partial\overline{\mathcal{C}}$ and $x\neq (0,0,0)$ we
have
\begin{equation} 
f^{(\nu )}(x+\mathbf{K})=\e^{i\theta K}f^{(\nu )}(x)\,, \quad \nu =0,1\,,
\end{equation}
where $f^{(0)}:=f,\,f^{(1)}:=\partial _{x_{1}}f.$ Now we define
$H_{\alpha, \Gamma} (\theta )$ as the self-adjoint Laplace
operator in $L^{2} (\mathcal{C})$ with the boundary conditions
introduced above; more precisely, $H_{\alpha }(\theta )$ is the
closure of
\begin{eqnarray*}
\dot{H}_{\alpha, \Gamma}(\theta ):D(\dot{H}_{\alpha,
\Gamma}(\theta )) &\!=\!& \{f\in \Upsilon_{\alpha }(\theta
):\dot{H}_{\alpha, \Gamma}(\theta ) f\in
L^{2}(\mathcal{C})\}\rightarrow L^{2}(\mathcal{C})\,, \\
\dot{H}_{\alpha, \Gamma}(\theta )f(x) &\!=\!& -\Delta f(x)\,,\quad
x\in \mathcal{C} \setminus \Gamma _{\mathcal{C}}\,.
\end{eqnarray*}
The following lemma states the usual unitary equivalence between
$H_{\alpha, \Gamma}$ and the direct integral of its fiber
components $H_{\alpha, \Gamma}(\theta ).$
\begin{lemma} \label{uniequ}
$\;UH_{\alpha, \Gamma}U^{-1}=\int_{[-\pi/K,\pi/K
)}^{\oplus}H_{\alpha, \Gamma}(\theta )\, d\theta$.
\end{lemma}
\begin{proof}
Take a function $f$ belonging to the set
\begin{equation} \label{definL}
\mathcal{L}:=\{g\in C^{2}(\R^{3}\backslash \Gamma )\,:\:
f\symbol{126}\alpha .bc(\Gamma )\,,\,
\mathrm{supp\,}f\;\mathrm{is} \; \mathrm{compact}\,\}\,
\end{equation}
then for all $i=1,2,3$ we have
$$
(U\partial _{i}f(x))_{\theta }=\partial _{i}(Uf)_{\theta }
(x)\,,\quad x\notin \Gamma\,,
$$
and the same relations hold for the second derivatives. Thus to
prove the lemma it suffices to show that any function admitting
the representation $(Uf)_{\theta }$ with $f\in \mathcal{L}$
belongs to $\Upsilon_{\alpha }(\theta )$. It is easy to check that
for all $x\neq (0,0,0)$ such that $x$ and $x+\mathbf{K}$ are in
$\partial\overline{\mathcal{C}}$ we have
$$
((Uf)^{(\nu )}_{\theta }(x+\mathbf{K})= \e^{i\theta K} ((Uf)^{(\nu
)}_{\theta}(x))\quad \mathrm{for} \quad \nu =0,1\,.
$$
The behavior of the function $(Uf)_{\theta }$ in the vicinity of
$\Gamma _{\mathcal {C}}$ is characterized by the limits $\Xi
((Uf)_{\theta })(\cdot)$ and $\Omega ((Uf)_{\theta })(\cdot)$.
Using the periodicity of $\Gamma $ we get
\begin{eqnarray*}
\Xi ((Uf)_{\theta })(s) &\!=\!& (2\pi)^{-1/2} \sum_{n\in \Z}
\e^{-in\theta K}\Xi (f)
(s+nL)\,,\quad s\in (0,L)\,, \\
\Omega ((Uf)_{\theta })(s) &\!=\!& (2\pi)^{-1/2} \sum_{n\in \Z}
\e^{-in\theta K}\Omega (f)(s+nL)\,,\quad s\in (0,L)\,;
\end{eqnarray*}
to derive these relations we used also the uniform convergence of
the sums. In this way we conclude that $(Uf)_{\theta }\symbol
{126}\alpha.bc(\Gamma _{\mathcal{C}}).$ The Laplace operator in
$L^{2}(\mathcal{C})$ with the domain consisting of functions which
admit the representation $(Uf)_{\theta }$ with $f\in \mathcal{L}$
is essentially self-adjoint and its closure coincides with
$H_{\alpha, \Gamma}(\theta)$; this completes the proof.
\end{proof}

\subsection{Spectral analysis of $H_{\protect{\alpha, \Gamma}}(\protect\theta )$}

As in the case of a finite curve we can now analyze the discrete
spectrum of the operator $H_{\alpha, \Gamma}(\theta )$. Before
doing that let us localize the essential spectrum. An argument
analogous to that of Proposition~\ref{essloop} shows that the
singular perturbation supported by $\Gamma_{\mathcal{C}}$ does not
change the essential spectrum of the Laplacian in a slab with
Floquet boundary conditions, i.e.
\begin{equation} \label{staess}
\sigma _{\mathrm{ess}}(H_{\alpha, \Gamma}(\theta ))=\left\lbrack
\theta ^{2},\infty \right)\,.
\end{equation}
To describe the asymptotic behavior of the eigenvalues of
$H_{\alpha, \Gamma}(\theta )$ we introduce a comparison operator
by $S_{\theta }=-\frac{d^{2}} {ds^{2}}-\frac{\kappa
(s)^{2}}{4}:\:D(S_{\theta })\to L^{2}(0,L)$, where
$$
D(S_{\theta }):=\{\, f\in W^{2,2}(0,L):\, f(L)=\e^{i\theta K
}f(0),\,f^{\prime }(L) =\e^{i\theta K}f^{\prime }(0)\,\}\,.
$$
In analogy with Theorem \ref{evloop} we state:
\begin{theorem} \label{evfibe}
Under the assumption given above for a fixed number n there exists
$\alpha (n)\in \R$ such that $\sharp \sigma _{d}(H_{\alpha
}(\theta ))\ge n$ holds for $\alpha \leq \alpha (n)$. Moreover,
the $j$-th eigenvalue of $H_{\alpha, \Gamma}(\theta)$ has the
asymptotic expansion of the form
$$
\lambda _{j}(\alpha ,\theta )=\xi _{\alpha }+\mu _{j}(\theta )+
\mathcal{O}(\e^{\pi \alpha })\quad \mathrm{as} \quad \alpha
\rightarrow -\infty\,,
$$
where $\mu _{j}(\theta )$ is the $j$-th eigenvalue of $S_{\theta
}$ and the error term is uniform with respect to $\theta $.
\end{theorem}
\begin{proof}
The argument follows closely that of Theorem \ref{evloop}; the
only difference is the replacement of periodic boundary condition
by the Floquet one. The fact that the error is uniform w.r.t.
$\theta$ is a consequence of Lemma \ref{eivalj} and continuity of
the functions $\mu _{j}(\cdot).$
\end{proof}

\subsection{Spectral analysis of $H_{\protect{\alpha, \Gamma}}$ in
terms of $H_{ \protect{\alpha, \Gamma}}(\protect\theta )$}

Now our aim is to express the spectrum of $H_{\alpha, \Gamma}$ in
the terms of $H_{\alpha, \Gamma}(\theta)$. First, let us note that
combining (\ref{staess}) with standard results
\cite[Sec.~XIII.16]{RS} we get the following equivalence for the
positive part of spectrum
$$
\sigma (H_{\alpha, \Gamma})\cap \lbrack 0,\infty )= \bigcup _
{\theta \in [-\pi/K,\pi/K )} \sigma (H_{\alpha, \Gamma}\,(\theta
))\cap \lbrack 0,\infty )=[0,\infty ).
$$
The negative part of spectrum is more interesting being given by
the union of ranges of the functions $\lambda _{j}(\alpha,\cdot)$.
They give rise to well-defined spectral bands because the latter
are continuous in the Brillouin zone $[-\pi/K,\pi/K )$. This can
be seen by checking in the usual way, putting $\theta$ into the
operator and showing that the $\theta$ dependent part is an
analytic perturbation. Alternatively, one can take $g=(Uf)_{\theta
}$ with $f\in \mathcal{L}$ as defined by (\ref{definL}) and
investigate the functions
$$
\theta \mapsto q_{g}(\theta ):=(g,H_{\alpha, \Gamma}(\theta )g)
_{L^{2}(\mathcal{C})} =\frac{1}{2\pi}\sum _{n, m\in
\Z}\e^{-i(n-m)\theta }(f_n,H_{\alpha,
\Gamma}f_{m})_{L^{2}(\mathcal{C})}\,,
$$
where $f_{n}(x):=f(x+n\mathbf{K})$. In view of (\ref{operaU}) and
the uniform convergence of the respective sums such a
$q_{g}(\cdot)$ is continuous for $g$ runing over a common core of
all $H_{\alpha, \Gamma}(\theta )$. Thus by the minimax priciple we
get the continuity of $\lambda _{j}(\alpha, \cdot)$ and combining
this fact with the results of \cite{RS} we get
$$\sigma (H_{\alpha, \Gamma})\cap (-\infty,0\rbrack =\left(
\bigcup _{\theta \in [-\pi/K,\pi/K )} \sigma (H_{\alpha }(\theta
))\right ) \cap (-\infty,0\rbrack $$
arriving finnaly at
$$
\sigma (H_{\alpha, \Gamma })=\bigcup _{\theta \in [-\pi/K,\pi/K )}
\sigma (H_{\alpha, \Gamma }(\theta ))\,.
$$
These results together with Theorem~\ref{evfibe} allow to describe
the band structure of $H_{\alpha, \Gamma}$, in particular, the
existence of gaps. Notice that this operator as well as
$S=-\frac{d^{2}}{ds^{2}}- \frac{\kappa (s)^{2}}{4} $ in
$L^{2}(\R)$ commute with the complex conjugation, so their Floquet
eigenvalues are generically twice degenerate depending on
$|\theta|$ only. For the comparison operator thus width of the
$j-$th gap is
\begin{eqnarray*}
G_{j}(S) =\mu_{j+1}(\pi/K)-\mu_{j}(\pi/K) \quad &\mathrm{
for}&\!\!\mathrm{odd}\; j \\
\mu _{j+1}(0)-\mu _{j}(0) \quad &\mathrm{ for}&\!\!
\mathrm{even}\; j
\end{eqnarray*}
and similarly for $H_{\alpha, \Gamma}$. The expansion of
Theorem~\ref{evfibe} then gives
$$
G_{j}(H_{\alpha, \Gamma}) =G_{j}(S)+\mathcal{O}(\,e^{\pi \alpha
})\,.
$$
In combination with the known result about existence of gaps for
one-dimen\-si\-onal Schr\"odinger operators we arrive at the
following conclusion.

\begin{corollary}
Suppose that in addition to the above assumption the function
$\kappa (\cdot)$ is nonconstant. In the generical case when $S$
has infinitely many open gaps, one can find to any $n\in\N$ an
$\alpha(n)\in\R$ such that the operator $H_{\alpha, \Gamma}$ has
at least $n$ open gaps in its spectrum if $\alpha< \alpha(n)$.
If the number of gaps in $\sigma(S)$ is $N<\infty$, then
$\sigma(H_{\alpha, \Gamma})$ has the same property for $-\alpha$
large enough.
\end{corollary}

\noindent Notice that this property is determined by the curvature
alone. Thus the result does not apply not only to the trivial case
of a straight line, but also to screw-shaped spirals $\Gamma$ for
which $\kappa$ is nonzero but constant.

\begin{remark} \label{crochet}
{\rm It is not always possible to choose $\mathcal{C}$ in the form
of a rectangular slab (\ref{percell}) as we did above, which would
satisfy the assumption (a$\Gamma _\mathrm{per}$); counterexamples
can be easily found. However, if we choose instead another period
cell $\mathcal{C}$ with a smooth boundary for which the property
(a$\Gamma _\mathrm{per}$) is valid, the argument modifies easily
and the claim of Theorem~\ref{evfibe} remains valid. On the other
hand, such a decomposition may not exist if the topology of
$\Gamma$ is non-trivial; a simple counterexample is given by a
``crotchet-shaped'' curve. While we conjecture that the claim of
Theorem~\ref{evfibe} is still true in this situation, a different
method is required to demonstrate it. }
\end{remark}


\subsection{Compactly disconnected periodic curves}

So far we have considered a single periodic connected curve. A
slightly stronger result about the existence of gaps in spectrum
of $H_{\alpha ,\Gamma }$ as $\alpha \to -\infty$ can be obtained
for compactly disconnected periodic curves in $\R^{3}$, i.e. such
that they decompose into a disjoint union in which each of the
connected components is compact. To be more specific, we consider
a family of curves obtained by translations of a loop $\Gamma
_{0}$ (being a graph of a function $\gamma _{0}$) generated by an
$r$-tuple $\{\mathbf{K}_{i}\}$ linearly independent vectors, where
$r=1,2,3$. The curve $\Gamma $ in question is  then a union
$\Gamma =\bigcup _{n\in \Z^{r}}\Gamma _{n}$, where $\Gamma _{n}$
are graphs of
$$
\gamma _{n}:=\gamma _{0}+\sum_{n\in \Z^{r}}
n_{i}\mathbf{K}_{i}:\:[0,L]\to \R ^{3}\,, \quad n=\{n_{i}\}\,;
$$
for the sake of brevity we put here $\Gamma _{n_{0}}=\Gamma
_{0},\: \gamma _{n_{0}}=\gamma _{0}$, where $n_{0}:=(0,0,0)$. We
assume that $\Gamma_0$ \emph{is contained in the interior} of the
period cell
$$
\mathcal{C}=\left\{\sum _{i=0}^{r-1}t_{i}\mathbf{K}_{i}:0\leq
t_{i}<1 \right\}\times \{\mathbf{K}_{i} \}^{\bot}\,,
$$
which is noncompact if $r=1,2$ and compact otherwise. Similarly as
before we can make Floquet-Bloch decomposition of $H_{\alpha
,\Gamma  }$ into a direct integral of the fiber operators
$H_{\alpha, \Gamma}(\theta )$. However now, since $\mathrm{dist}
(\partial \mathcal{C}, \Gamma_{\mathcal{C}})>0$ holds by
assumption, the comparison operator $S=S(\theta)$ is now
independent of the quasimomentum $\theta \in \prod_{1\le i\le
r}^\times [-\pi| \mathbf{K}_{i}|^{-1}, \pi|\mathbf{K}_{i}|^{-1})$.
While in the previous case some gaps of $S(\theta)$ might be
closed, now they are all open. As a result each gap in the
spectrum of $\sigma (H_{\alpha ,\Gamma })$, which depends of
course on $\theta$, will eventually open for $-\alpha$ large
enough.
\begin{theorem}
Under the assumptions stated above the spectrum of $H_{\alpha
,\Gamma }(\theta )$ is purely discrete if $r=3$, and $\sigma
_{\mathrm{ess}}(H_{\alpha ,\Gamma }(\theta))= \left\lbrack\,
\sum_{i=1}^r \theta_i^2,\infty \right)$ if $r=1,2$. The $j$-th
eigenvalue of $H_{\alpha ,\Gamma }(\theta )$ admits the asymptotic
expansion of the following form,
$$
\lambda _{j}(\alpha ,\theta )=\xi _{\alpha}+\mu
_{j}+\mathcal{O}(\e^{\pi \alpha })\quad \mathrm{as} \quad \alpha
\to -\infty\,,
$$
where $\mu_j$ is the $j$-th eigenvalue of $S$ and the error is
uniform w.r.t. $\theta$. Consequently, for any $n\in\N$ there is
$\alpha(n)\in \R$ such that the operator $H_{\alpha, \Gamma}$ has
at least $n$ open gaps in its spectrum if $\alpha< \alpha(n)$.
\end{theorem}


\section{Concluding remarks}

(a) The results obtained in the previous discussion can be
rephrased as a \emph{semiclassical approximation.} To see this let
us consider the Hamiltonian $H_{\alpha ,\Gamma }(h)$ with the
Planck's constant $h$ reintroduced; the latter is understood in
the mathematical sense, i.e. as a parameter which allows us to
investigate the asymptotic behavior as $h\to 0$. The operator in
question then acts as
$$ H_{\alpha ,\Gamma }(h) f(x)=-h^{2}\Delta f(x)\,,\quad x\in
\R^{3}\setminus \Gamma\,, $$
and has the domain
$$D(H_{\alpha ,\Gamma }(h) )=\{f\in \Upsilon
_{\R^{3}}:f\symbol{126}\alpha (h).bc (\Gamma )\}\,, $$
where
\begin{equation} \label{semialpha}
\alpha (h):=\alpha +\frac{1}{2\pi }\ln h\,.
\end{equation}
This definition of $H_{\alpha ,\Gamma }(h)$ requires a comment. In
the case $\mathrm{codim\,}\Gamma=1$ discussed in \cite{Ex} the
Hamiltonian is defined by the natural quadratic form, hence
introducing $h$ means a multiplicative change of the coupling
parameter, $\alpha \to \alpha h^{-2}$; one can see that also from
the approximation of such an operator by means of scaled regular
potentials \cite{EI}.

In contrast to that a two-dimensional point interaction involves a
complicated nonlinear coupling constant renormalization
\cite[Sec.~I.5]{AGHH}, so introducing Planck's constant is in this
case arbitrary to a certain extent. We choose the simplest way
noticing that the relation between the free operators $-\Delta$
and $-h^2\Delta$ can be expressed by means of the scaling
transformation $x\mapsto hx$, and require the similar behavior for
the singular interaction term; it is well known that a scaling for
a two-dimensional point interaction is equivalent to a logarithmic
shift of the coupling parameter -- cf.~\cite{EGST}. In view of
(\ref{semialpha}) the semiclassical limit $h\to 0$ is within this
convention for a fixed coupling constant $\alpha $ equivalent to
$\alpha (h)\to -\infty$ which means a strong coupling again. Since
$H_{\alpha ,\Gamma}(h) = h^2 H_{\alpha(h) ,\Gamma }(1)$ we see
that the eigenvalues $\lambda_{j }(\alpha, h)$ of $H_{\alpha
,\Gamma }(h)$ take then the following form,
$$ \lambda_{j }(\alpha, h)=\xi_{\alpha}+\mu_{j}h^{2}
+\mathcal{O}(h^{5/2})\quad \mathrm{as} \quad h\to 0\,. $$
In the same way we find the counting function which is given by
$$
\#\sigma _{d}(H_{\alpha }(h))=\frac{L}{\pi h}(-\xi
_{\alpha})^{1/2}(1+\mathcal{O} (h^{1/4}))\,.
$$

\vspace{.5em}

\noindent (b) Let us finally list some \emph{open problems}
related to the present subject:
\begin{itemize}
\item One is naturally interested in the asymptotic expansion in
the situation when $\Gamma$ is a curve with free ends and the
present method allows us to treat the counting function only; the
analogous question stands for planar curves \cite{EY1} and
surfaces with a boundary \cite{Ex}. We \emph{conjecture} that the
expansion of Theorem~\ref{evloop} holds again with $\mu_j$
corresponding to the comparison operator which acts according to
(\ref{compar}) with \emph{Dirichlet} boundary conditions at the
boundary of $\Gamma$.
\item The results can be extended to higher dimensions provided
$\mathrm{codim\,}\Gamma\le 3$ so that the singular interaction
Hamiltonian is well defined.
\item The smoothness assumption is crucial in our argument. A
self-similar curve such as a broken line consisting of two
halflines joined at a point provides an example of a situation
where the asymptotic behavior differs from that of
Theorem~\ref{evloop}. One can ask, e.g., how the asymptotics looks
like for a piecewise smooth curve with non-zero angles at a
discrete set of points.
\item Another important question concerns the \emph{absolute
continuity} of the spectrum in case when $\Gamma$ is a periodic
curve or a family of curves. The answer is known if
$\mathrm{codim\,}\Gamma=1$ and the elementary cell is compact
\cite{BSS, SuS}. The cases of a single connected periodic curve or
a periodic surface diffeomorphic to the plane are open, and the
same is true for periodic curve(s) in $\mathbb{R}^3$, i.e. the
situation with $\mathrm{codim\,}\Gamma=2$.
\end{itemize}


\subsection*{Acknowledgment}

We thank the referees for their remarks which helped to improve
the text. S.K. is grateful for the hospitality in the Department
of Theoretical Physics, NPI, Czech Academy of Sciences, where a
part of this work was done. The research has been partially
supported by the ASCR project K1010104 and by the Polish Ministry
of Scientific Research and Information Technology under
(solicited) grant No PBZ-Min-008/PO3/03.


\end{document}